\documentstyle[11pt]{article}

\def\e{\epsilon}

\def\fr{\frac}

\def\t{\tau}

\def\del{\partial}

\let\bm=\bibitem

\newcommand{\vs}[1]{\vspace{#1 mm}}
\newcommand{\hs}[1]{\hspace{#1 mm}}

\def\be{\begin{equation}}
\def\ee{\end{equation}}
\def\bea{\begin{eqnarray}}
\def\eea{\end{eqnarray}}
\def\bar{\begin{array}}
\def\ba{\begin{array}}

\newcommand{\tamphys}{\it Center for Theoretical Physics, Texas
A\&M University, College Station, TX 77843, USA}

\newcommand{\auth}{\large
N.S. Deger\footnote{sadik@rainbow.physics.tamu.edu} and
A. Kaya\footnote{ali@rainbow.physics.tamu.edu}}

\begin{document}

\baselineskip=.55cm


\hfill{hep-th/0010141}

\hfill{\today}

\vs{20}

\begin{center}

{\Large \bf $AdS$/CFT and Randall-Sundrum Model Without a Brane}
\\ 

\vs{15}

\auth \\

\vs{5}

\tamphys

\vs{30}

{\bf Abstract}

\end{center}

We reformulate the Randall-Sundrum (RS) model on the compactified $AdS$ by
adding a term proportional to the area of the boundary to the usual
gravity action with a negative cosmological constant and show that gravity
can still be localized on the boundary without introducing singular brane
sources. The boundary conditions now follow from the field equations,
which are obtained by letting the induced metric vary on the boundary.
This approach gives similar modes that are obtained in \cite{sun} and
clarifies the complementarity of the RS and the $AdS$/CFT pictures.
Normalizability of these modes is checked by an inner-product in the space
of linearized perturbations. The same conclusions hold for a massless
scalar field in the bulk.
\vs{20}


\pagebreak 

\setcounter{page}{1}



 
Any realistic theory of gravity should be able to produce
$r^{-1}$ behavior of the gravitational potential. Generically,
the potential falls off like $r^{-d_{inf}+3}$, where $d_{inf}$ is
the number of dimensions with infinite extend. Thus, to get
$r^{-1}$ behavior, higher dimensional theories of gravity have
been assuming compactness of the extra dimensions. However, last
year Randall and Sundrum \cite{sun} showed that this result can
still be obtained allowing a non-compact fifth dimension. Their
now well-known model (RS)  consists of a positive tension 3-brane
in $AdS_5$ which corresponds to our universe and it was shown in
\cite{sun} that gravity is localized on it with the usual
Newtonian behavior.

\
\

It is reasonable to try to apply the $AdS$/CFT duality \cite{mal, gub,
wit} to understand this phenomena. This has already been pointed out by
Maldacena and Witten in their unpublished remarks and has been pursued in
several papers \cite{ver1, gubser, lin, haw, duff, ver2, brett}. In these,
the brane is thought to be located at finite radial distance which can be
considered as the boundary of the $AdS$ space. Geometrically, this gives
rise to a compact slice of the $AdS$ space. $AdS$/CFT correspondence now
implies that RS model is equivalent to a four dimensional gravity coupled
to a strongly interacting CFT. \footnote{For applications of this
equivalence  to cosmology see, for instance, \cite{cos1,cos2, cos3,
cos4}.}

\ 
\ 

However, there is an apparent difficulty in the above considerations. On
the RS side, the presence of a singular brane in the $bulk$ is very
crucial \cite{sun,lin}. Indeed, without the brane one has two massless
modes and two towers of continuum massive modes on $AdS$. Introducing the
brane, the field equations pick up a delta function source and RS
background now becomes a solution. \footnote{Several no-go theorems have
been proved \cite{nogo1, nogo2, nogo3, nogo4, nogo5} stating that there
are no $smooth$ RS compactifications for a large class of supergravity
theories.} However, once the brane is assumed to be on the boundary, then
this delta function is lost since the brane action does not anymore modify
field equations which are derived by keeping the metric fixed on the
boundary.  Therefore, in this case it is not obvious how to repeat the
calculations of \cite{sun,lin}.  On the other hand, on the $AdS$/CFT side,
the role played by this dynamical brane is not clear. For instance, it is
not known how to include the degrees of freedom associated with it in the
path integral.

\
\

In this letter we will show how these problems can be solved by replacing
the brane action in the bulk with a term proportional to the area of the
boundary of the compactified AdS space. It turns out, in deriving field
equations if one assumes that the metric on the boundary is not fixed,
then this gives a boundary condition. \footnote{See also \cite{ak} for a 
different variational approach.} Indeed, this is not an assumption
but a necessity when in the path integral quantization one includes the
degrees of freedom associated with the boundary metric. After obtaining
the boundary condition, one finds a single massless mode and a tower of
continuum massive Kaluza-Klein (KK) modes on $AdS$. As it will be shown,
the massless mode which extends from horizon to the boundary at infinity 
is not normalizable. However, removing the asymptotic region gives rise to
normalizable modes. Thus the RS scenario can be reformulated on a
compactified AdS space, which may arise as part of the string
compactifications on orientifold and/or orbifold spaces, as discussed in
\cite{ver2}. With this understanding, it is not necessary to assume
the existence of a singular, dynamical brane (for instance located at the
boundary).

\
\

Assuming that such a model arises in a string/M theory
compactification, it is natural to consider presence of different fields
in the bulk other than gravity. However, in the brane world scenario it is
not very clear how to treat, for instance, a scalar field since there is
no unique and natural coupling of a scalar to the brane. Another
advantage of the approach presented in this letter is that, other fields
propagating in the bulk can be treated exactly like gravity; one starts
from the well-known action which is used in $AdS$/CFT, and obtains the
equations of motion by allowing field variations on the boundary.  To
demonstrate this we will consider a minimally coupled scalar field and
find, on the boundary, a single massless mode and a tower of heavier,
continuum KK modes. Including gravity, $AdS$/CFT implies a dual
description which corresponds to gravity and a scalar field coupled to a
strongly interacting CFT on the boundary. In one picture, the graviton and
scalar field propagators pick up corrections due to the exchange of KK
modes and in the dual picture these corrections arise from the coupling of
fields to CFT. Specifically, two-point function of the energy-momentum
tensor modifies graviton propagator. On the other hand, the two-point
function of the operator in the CFT, which is dual to the scalar at hand
in AdS/CFT, gives rise to a correction for the scalar propagator. We
clarify the passage from the RS picture to $AdS$/CFT description by
evaluating the same partition function using either $AdS$/CFT or a
semiclassical approximation.

\
\

In $d+1$-dimensions, field equations for gravity with a negative
cosmological constant $\Lambda$ are
\be\label{eq1}
R_{AB} -\fr{1}{2}G_{AB}R=\Lambda G_{AB},
\ee
which admit the AdS space
\be \label{met}
ds^{2}= \fr{l^{2}}{z^{2}}(dz^{2}+ \eta_{\mu\nu}dx^{\mu}dx^{\nu})
\ee
as a solution with $l^{2}= - d(d-1)/(2\Lambda )$. Identifying $z$
as the radial coordinate, AdS space can be viewed to be the
metric of a domain wall spanned by the coordinates $x^{\mu}$. The
domain wall will serve as a model for our observed universe and
at this moment can be thought to be located at an arbitrary
radial position. Let us now discuss if such a scenario can
produce the well known properties of gravity without a
cosmological constant. We first note that in the absence of
matter, the metric on the domain wall is flat. Assuming existence
of matter on the wall, one should replace the flat metric
$\eta_{\mu\nu}$ in (\ref{met}) with a curved one $g_{\mu\nu}(x)$.
Then, in vacuum, (\ref{eq1}) implies the Ricci flatness of
$g_{\mu\nu}$. Thus, it seems that one can easily recover the two
basic properties of gravity without a cosmological constant.

\
\

Let us now consider a similar calculation from an effective action point
of view. In $d+1$-dimensions the action is, 
\be\label{act1} 
S=\fr{1}{16\pi G_{d+1}}\int_{M} \hs{2} \sqrt{-G} \hs{2}(R-2\Lambda )\hs{2}
+ \fr{1}{8\pi G_{d+1}} \int_{\partial M} \hs{2} \sqrt{-\gamma} \hs{2}K ,
\ee 
where $G_{d+1}$ is the Newton's constant and $\gamma_{\mu\nu}$ is the
induced metric on the boundary. Since, in deriving field equations
(\ref{eq1}) the metric is held fixed on the boundary, there is a certain
freedom of adding boundary terms to the action. The most natural
modification is to add
\be
\label{act2} S_{1}= \fr{a}{16\pi G_{d+1}} \int_{\partial M}
\sqrt{-\gamma}, 
\ee 
where $a$ is a constant. This term is proportional to the area of the
boundary and its presence was first discussed in \cite{liu}. It is worth
to mention that the area term has been used in the AdS/CFT context for
some time. Following \cite{liu}, the same term was present in the
calculation of Weyl anomaly in the paper \cite{ek1}. This term was
motivated further by Hamiltonian formulation of the supergravity action in
\cite{ek2}. And finally, it was found by using the counterterm technique
in several papers including \cite{ek3}. Let us emphasize that in these
papers this term was $not$ considered as a brane source as in our approach
in this letter.

\
\

For the metric, we will assume
\be \label{met2} ds^{2}= \fr{l^{2}}{z^{2}}(dz^{2}+
g_{\mu\nu}(x)dx^{\mu}dx^{\nu}), 
\ee 
and try to obtain an effective
action for $g_{\mu\nu}(x)$. Although the boundary is located at $z=0$, we
cut the asymptotic region at the surface $z=\e$ and name this space as
$AdS_{\e}$. Inserting (\ref{met2}) into the total action $S+S_{1}$ and
choosing 
\be\label{a} 
al= -2(d-1), 
\ee 
we obtain 
\be 
S+S_{1} =
\fr{1}{16\pi G_{d}} \int d^{d}x \sqrt{-g} \hs{2} {\cal{R}}, 
\ee 
where
$\cal{R}$ is the Ricci tensor of $g_{\mu\nu}$ and the $d$-dimensional
Newton's constant is given by 
\footnote{Let us note that the same $\epsilon$ dependence in (8) has also
been obtained in RG flow considerations, see for example \cite{rg}}
\be\label{eq2} 
G_{d}= (d-2)\fr{\e^{d-2}}{l^{d-1}}G_{d+1}. 
\ee 
Thus, to get a proper effective action describing pure gravity without
a cosmological constant, one should start with the total action $S+S_{1}$
where the free parameter $a$ in $S_{1}$ is fixed by (\ref{a}). Although,
this exactly corresponds to the fine tuning in the original RS model,
there is an important difference. In the RS model, $S_{1}$ has been
replaced with a singular brane action $S=\t \int_{\Sigma} \sqrt{-\gamma}$
coupled to the gravity, where $\Sigma$ is the world-volume embedded in the
bulk. To cancel the contribution of the bulk cosmological constant, the
tension $\t$ of the brane should be fine tuned. Thus, the cosmological
constant problem is pushed into the brane. However, in the above case, one
may expect $S_{1}$ with coefficient (\ref{a}) to come from an $AdS$
compactification of string/M theory. Indeed, let us note that $exactly$
$the$ $same$ term should be added to (\ref{act1}) to obtain conformally
invariant graviton 2-point function using $AdS$/CFT duality \cite{liu},
which shows that in order for $AdS$/CFT duality to work, such a boundary
term should come from an $AdS$ compactification of string/M theory. This
raises the possibility of solving the cosmological constant problem not by
fine tuning but by the presence of a critical boundary action which may arise
after compactification.

\
\

From (\ref{eq2}), it is clear that $\e$ should be kept non-zero.
The strength of gravity in $d$-dimensions is determined by the
$d+1$-dimensional Plank scale, cosmological constant $\Lambda$
and a new length scale $\e$. Choosing $\e$ to be very small (and
$\e<l$), one can generate a very large Plank mass in
$d$-dimensions from a $d+1$-dimensional Planck mass which may be
of the order of standard model energy scales. This is similar to
the original proposal of \cite{rs2} to solve the hierarchy
problem and is related to the AdS geometry. Note that this problem cannot
be addressed in the second RS model \cite{sun}. At this point it is
natural to assume the domain wall to be located at $z=\e$. We
note that, we do not think of this wall as a dynamical object. It
simply means that we live at the boundary at $z=\e$ and by some
mechanism we cannot penetrate to the region $z>\e$, at least at
low energies.

\
\

Although, this scenario seems to work well in $d$-dimensions, let us
mention one potential $d+1$-dimensional problem. From (\ref{met2}), a
simple calculation gives \cite{hor}
\be\label{singu}
R_{ABCD}R^{ABCD}= \fr{2d(d+1)}{l^{4}} + \fr{z^{4}}{l^{4}}{\cal{R}_
{\mu\nu\sigma\rho}}{\cal{R}}^{\mu\nu\sigma\rho},
\ee
where $\cal{R}_{\mu\nu\rho\sigma}$ is the curvature tensor of
$g_{\mu\nu}$, which indicates a generic curvature
singularity on the ``horizon'' at $z=\infty$. To avoid such a
singularity, one may assume the existence of another wall located at
a finite $z$ value, which gives rise to a radial coordinate
compactified on a line. This is similar to the Horava-Witten
construction in 11-dimensions \cite{hw}, where one coordinate of
the flat Minkowski space has been compactified by two 10-branes.
In this paper, we do not attempt to modify the discussion by
introducing another domain wall. However, it is worth to note
that the singular term in (\ref{singu}) vanishes for special
backgrounds or may not arise when $g_{\mu\nu}$ in (\ref{met2})
has a non-trivial $z$ dependence. For instance, the gravitational
waves propagating along the $d$-dimensional boundary can be
described by plane wave backgrounds, and for such solutions all
curvature invariants, such as the singular term in (\ref{singu}),
vanish. Also, the strength of gravity due to a static source $on$
the boundary turns out to fall down as $z\to\infty$ \cite{lin},
and there is no singularity problem for this case.

\
\

After these effective theory considerations, let us now determine the
spectrum of modes on $AdS_{\e}$. To include the degrees of freedom
associated with the gravity on the boundary, one needs to let the
induced metric to vary in deriving the field equations. The boundary
itself can still be thought fixed, which implies $\delta n_{A}=0$ and
$\delta n^{A}=0$ $on$ $\partial M$, where $n_{A}$ is the unit normal
vector to the boundary. \footnote{This condition can also be imposed as a
gauge choice on the boundary.} It is not obvious if after these
modifications the variation of $S+S_{1}$ is well defined or if $AdS_{\e}$
is still a solution. To verify this, we vary the action $S+S_{1}$ by
carefully treating the boundary terms coming from integration by parts,
and obtain 
\bea 
\delta (S+S_{1}) &=& \frac{1}{16\pi G_{d+1}} \int_{M}
\sqrt{-G} (R_{AB} -\fr{1}{2}G_{AB}R-\Lambda G_{AB})\delta G^{AB} \nonumber
\\ \label{variation}
 &+& \frac{1}{16\pi G_{d+1}} \int_{\partial M} \sqrt{-\gamma}(K_{AB}-\gamma_
{AB}K- \frac{a}{2} \gamma_{AB})\delta \gamma^{AB}=0,
\eea
where $\gamma_{AB}$ is the induced metric on $\partial M$, $K_{AB}$ is the
extrinsic curvature of the boundary defined by
$K_{AB}=\gamma_{A}{}^{C}\gamma_{B}{}^{D}\nabla _{C}n_{D}$ and
$K=\gamma^{AB}K_{AB}$. In deriving this result, we introduced an adapted
coordinate system (Gaussian normal coordinates) to the boundary so that
the metric near $\partial M$ can be written as
\be
ds^{2}= \frac{l^{2}}{z^{2}}dz^{2} + \gamma_{\mu\nu}(x,z)dx^{\mu}dx^{\nu}.
\ee
The boundary is located at $z=\e$ and $n=-(z/l)\partial_{z}$ is
the unit normal vector. In this coordinate system, $\delta
G_{AB}\to \delta \gamma_{\mu\nu}$ on $\partial M$ (since $\delta
n^{A}=0$ and $\delta n_{A}=0$ imply $\delta G_{AB}n^{B}=0$ and
$\delta G^{AB}n_{B}=0$ on $\partial M$) and the extrinsic
curvature is given by
$K_{\mu\nu}=-\e/(2l)\partial_{z}\gamma_{\mu\nu}|_{z=\e}$.  Field
equations following from (\ref{variation}) reads
\bea 
R_{AB} -\fr{1}{2}G_{AB}R-\Lambda G_{AB} & = & 0, 
\label{neq1}
\\ (K_{AB}-\gamma_{AB}K - \frac{a}{2}
\gamma_{AB})|_{\partial M} & = & 0. 
\label{neq2} 
\eea 
Although the bulk equations remain intact, we obtain a boundary
condition for the metric. It turns out that $AdS_{\e}$ obeys
(\ref{neq2}) only if the free parameter $a$ is fixed by
(\ref{a}). Remarkably, the same fine tuning, which has been
imposed to cancel the induced cosmological constant on the
boundary, is now required to have $AdS_{\e}$ as a solution of the
theory.

\
\
 
Assuming small perturbations around $AdS_{\e}$, the metric of the
wall in (\ref{met2}) can be written as
$g_{\mu\nu}=\eta_{\mu\nu}+h_{\mu\nu}(x,z)$. Imposing the gauges
$\del_{\mu}h^{\mu}{}_{\nu}=0$ and $h^{\mu}{}_{\mu}=0$, the
linearized equations following from (\ref{neq1}) and (\ref{neq2})
read 
\be\label{diff} z^{2} \del_{z}^{2}\hat{h}_{\mu\nu} -
(d-1)z\del_{z}\hat{h}_{\mu\nu} + m^{2} z^{2} \hat{h}_{\mu\nu} =
0, 
\ee 
\be
\label{bound} 
\partial_{z}\hat{h}_{\mu\nu}|_{z=\e}=0,
\ee 
where $h_{\mu\nu}= e^{ip.x}\hat{h}_{\mu\nu}(z)$ and $m$ is
the $d$-dimensional mass given by $m^{2}=-p^{2}$. The solutions
of (\ref{diff}) obeying the boundary condition (\ref{bound}) are
\be\label{soln} 
\hat{h}= \begin{cases}{ \textrm{const.}; \hs{5}
\textrm{when}\hs{2}m=0, \cr\cr N_{m}(z/l)^{d/2}
(A_{m}J_{d/2}(mz)+B_{m}Y_{d/2}(mz)); \hs{2}
\textrm{when}\hs{2}m\not = 0, } \end{cases} 
\ee 
where $N_{m}$ is a normalization constant to be fixed in a moment, and
$A_{m}=Y_{d/2-1}(m\e )$, $B_{m}=-J_{d/2-1}(m\e )$. For $m=0$,
$\hat{h}=z^{d}$ also solves (\ref{diff}), but does not obey the boundary
condition (\ref{bound}). (\ref{soln}) is very similar to the modes found
in \cite{sun} where, due to the presence of a singular brane, there is an
extra delta function source in (\ref{diff}). This implies a boundary
condition along the brane eliminating one of the possible zero modes
(corresponding to $\hat{h}=z^{d}$ mode). We see that a similar boundary
condition can be obtained from field equations by allowing the induced
metric on the boundary to vary, giving again a single massless mode and a
tower of continuum massive KK modes in $d$-dimensions.

\
\

At this point, one should make sure of the regularity and the
normalizability of these perturbations, which are of the form
\be\label{form}
\delta G_{zz}=0, \hs{4} \delta G_{z\mu}=0,\hs{4} \delta G_{\mu\nu}= 
\frac{l^2}{z^{2}}h_{\mu\nu},
\ee
when viewed in $d+1$-dimensional space. Since we remove the
asymptotic region, horizon is the only possible location for
irregularity. Note that, the Bessel function $Y$ diverges at
$z=0$, therefore the massive KK modes are irregular on $AdS$
boundary when $\e=0$. The mode $\hat{h}=\textrm{const.}$
represents gravitational waves along the boundary and are regular
as discussed below (\ref{singu}). Unfortunately, KK modes seem to
be irregular at the horizon. To see this we construct the scalar
$\delta G_{AB}\delta G_{CD}G^{AC}G^{BD}$, which diverges like
$z^{d-1}$ as $z\to \infty$. This indicates that, contrary to the
waves propagating along the boundary, the ones moving through the
horizon may be irregular. Let us note that the same type of
singularity also arises in the RS model \cite{sun}. Whether a divergence of
this type has a physical significance is not very clear to us.
One may still consider a perturbation of this type to be regular,
if the tidal forces on geodesics and all curvature invariants of
the perturbed metric turn out to be finite. 

\
\

It is also interesting to consider the singularity problem in Euclidean
signature. The massive modes on Euclidean $AdS_{\e}$ can be obtained from
(\ref{soln}) by analytical continuation $m \to im$, which become a linear
combination of the modified Bessel functions $K_{d/2}(mz)$ and
$I_{d/2}(mz)$. Although $K$ decays as $z\to\infty$, $I$ diverges
exponentially, which shows that the massive modes are singular in
Euclidean signature. It is possible to overcome this problem by placing
another wall at $z=L$. This implies another boundary condition at $z=L$
and $m$ should have discrete values to satisfy this condition.  
\footnote{See \cite{br} for a discussion of the scalar field
quantization on cut-off $AdS$ at $z=L$.} As $L\to\infty$, one recovers the
continuum spectrum. Here in this letter, we do not consider the
consequences of this modification.

\
\

To check normalizability, one should introduce a suitable
inner-product in the space of linearized perturbations. Noting
that (\ref{diff}) is the Laplacian acting on the scalars, we
define
\be\label{inner}
<h , h'>=-i\int_{\Sigma}\hs{2} ( h\hs{1}\del_{\mu}\hs{1}h'^{*} - 
h'^{*}\hs{1}\del_{\mu}\hs{1} h ) \hs{1}n^{\mu}d\Sigma,
\ee
where $\Sigma$ is a $t= \textrm{constant}$ hypersurface,
$d\Sigma$ is the induced volume element and $n^{\mu}$ is the
future directed unit normal. As we will see, (\ref{inner}) does
not depend on the choice of $\Sigma$ for the modes (\ref{soln}).
Using the plane wave dependence of the wave functions along
$x$-directions, one finds
\be\label{inner2}
<h , h'>= (2\pi )^{d-1} \delta (\vec{p} -\vec{p}')(\omega + \omega ')
e^{i(\omega '-\omega )t} I(m,m'),
\ee
where $\omega \equiv p_{0}$,  and 
\be \label{I}
I(m ,m')= \int_{\e}^{\infty} \hs{1}dz \frac{l^{d-1}}{z^{d-1}} 
\hs{1}\hat{h}\hs{1}\hat{h'}.
\ee
From (\ref{I}), the massless mode turns out to be normalizable if
and only if $\e\not =0$. Note that, in this case
$\vec{p}=\vec{p}'$ implies $\omega=\omega '$ and the time
dependence in (\ref{inner2}) disappears. For massive modes, using
the Bessel's differential equation, we obtain
\be
I(m,m') = \frac{ N_{m}N_{m'}}{m^{2}-m'^{2}} 
\left[ \frac{z}{l} \left( f_{m} \frac{d}{dz}f_{m'} - f_{m'} \frac{d}{dz}f_{m}
\right) \right]^{\infty}_{\e},
\ee 
where $f_{m}(z)=( A_{m}J_{d/2}(mz)+B_{m}Y_{d/2}(mz))$. There is no
contribution from $z=\e$ due to the boundary condition obeyed by
$f_{m}(z)$ at $z=\e$. Evaluating $z=\infty$ limit using the asymptotic
forms of the Bessel functions, we find
\be
I(m,m')= \frac{\sqrt{2\pi}}{ml}(A_{m}^{2}+B_{m}^{2})\delta (m-m').
\ee
Therefore, choosing $N_{m}= (ml/[2\omega
(A_{m}^{2}+B_{m}^{2})])^{1/2}$, the massive modes are normalized
\be\label{KKprod}
<h,h'>= (2\pi )^{d-1}\sqrt{2\pi}\delta (\vec{p}-\vec{p}')\delta(m-m'),
\ee
which is again independent of $t$. The product of the massless
mode with a massive KK mode turns out to be zero since (\ref{I})
becomes
\bea
I(m,0)&\sim & \int_{\e}^{\infty} z\hs{1} dz \hs{1}f_{m}\hs{1} z^{-d/2}\\
&\sim& \frac{1}{m^{2}}\left[ z \left( f_{m} \frac{d}{dz}z^{-d/2} - z^{-d/2} 
\frac{d}{dz}f_{m} \right) \right]^{\infty}_{\e}=0,
\eea
where we have again used the Bessel's differential equation and
the boundary condition obeyed by $f_{m}(z)$ at $z=\e$. This shows
that all inner products are well defined and do not depend on
$t$. Since the set (\ref{soln}) turns out to be $complete$ and
$orthonormal$, a generic gravitational wave can be obtained by
superposing these modes and their complex conjugates. The
coefficients of positive and negative frequency modes, $a_{m}(p)$
and $a_{m}^{\dagger}(p)$, are interpreted as creation and
annihilation operators on a Fock space of particle states,
respectively. Since the background geometry is static, there is
no ambiguity in defining the vacuum. The particle created by
$a_{m}^{\dagger}(p)$ has the mass $m$ and momentum $p$ in
$d$-dimensions. We note that these are valid when $\e\not
=0$. On $AdS$, when $\e=0$, the inner products and the Fock space of states 
are not well defined indicating the decoupling of gravity.

\
\

From (\ref{KKprod}), we see that the measure on the set of
continuum eigenvalues is simply $dm$. Since the dimensionless
coordinate on $AdS_{\e}$ is $z/l$, the corresponding
dimensionless eigenvalues and the measure are $l\hs{1}m$ and
$ldm$, respectively. Following \cite{sun}, we can now answer why
KK modes below the accessible energy scale of current experiments
are not observed. To calculate the probability of creating such a
massive mode in a process, we should sum over the continuum
eigenvalues up to the energy scale of the process, which is
$\int^{p} \hs{1}(m\e)^{d-3}(\e /l)^{d-1} ldm$, where the factor
$(m\e)^{d-3}(\e /l)^{d-1}$ comes from the continuum wave function
suppression at $z=\e$. Thus, creation of a massive KK mode is
suppressed  by a factor of $(\e^{2} p/l)^{d-2}$. This is not
surprising, since the same conclusion holds for the RS
model and the modes found in this paper are similar to the ones
found in \cite{sun}.

\
\

The graviton propagator can be calculated either by direct
construction or by superposing the modes (\ref{soln}) since they
are complete. In the former case, following (\ref{diff}), one
needs to solve
\be\label{green}
\Box\Delta(x,z;x',z')=\frac{\delta^{d}(x-x')\delta(z-z')}{\sqrt{-G}},
\ee
\be\label{greenb}
\partial_{z}\Delta|_{z=\e}=0
\ee
where $\Box$ is the Laplacian and the boundary condition (\ref{greenb}) is
implied by (\ref{bound}). This Green function has been constructed in
\cite{lin}, and is localized near the boundary. Also, assuming the sources
to be located on the boundary, $\Delta$ can be separated into the standard
$d$-dimensional propagator plus a piece coming from the exchange of KK
modes \cite{lin}, which in momentum space reads
\be\label{cgreen}
\Delta (p)=\frac{d-2}{\e}\hs{1}\frac{1}{p^{2}}\hs{1}+
\hs{1}\Delta_{KK}(p),
\ee
where
\be\label{KK}
\Delta_{KK}=-\frac{1}{p}\hs{1}\frac{H^{(1)}_{d/2-2}(p\e)}{H^{(1)}_{d/2-1}(p\e)},
\ee
and $H^{(1)}$ is the first Hankel function defined by $H^{(1)}=J+iY$. 

\
\

In the path integral quantization of the system in a semiclassical
approximation, one may start from the following integral 
\be\label{firstint}
Z = \hs{1}\int [dG]\hs{1} e^{iS\hs{1}+\hs{1}iS_{1}},
\ee
where $S$ and $S_{1}$ are given in (\ref{act1}) and (\ref{act2}), and the
sum is over $all$ metrics on $M$ (no boundary condition imposed on the
induced metric on $\partial M$).\footnote{The unit normal vector of
$\partial M$ can be fixed by a gauge choice (normal coordinates adapted to
boundary). The remaining gauge freedom can be ignored for the following
discussion.} This integral can be evaluated using saddle point
approximation when there is an extremum of the functional $S+S_{1}$ under
these conditions. As we discussed above, such an extremum should obey
(\ref{neq1})-(\ref{neq2}), and $AdS_{\e}$ is a solution when $a$ is fixed
as in (\ref{a}). One can then determine the small fluctuations around
$AdS_{\e}$ like (\ref{soln}) and construct the propagator of the theory
obeying (\ref{green}) and (\ref{greenb}). In this approximation,
(\ref{firstint}) becomes proportional to the $(\textrm{det} \Delta)^{-1/2}$.

\
\

The path integral approach allows one to see the existence of a
complementary picture implied by $AdS$/CFT correspondence. For this, one
can evaluate (\ref{firstint}) by first summing over all bulk metrics which
match a given boundary metric, and then integrating over all boundary
metrics \cite{gubser,lin,haw}. This gives
\be\label{int1}
Z=\int \hs{1}[d\gamma]\hs{1} Z[\gamma],
\ee
where
\be\label{last}
Z[\gamma] = \hs{1} e^{iS_{1}[\gamma ]} \hs{1} \int [dG]_{G|_{\partial M}=
\gamma} \hs{1} e^{iS},
\ee
and $\gamma$ is the induced metric on the boundary. The integral
(\ref{last}) can be calculated by a version of AdS/CFT duality
for a finite radial coordinate which, after including
necessary counterterms required to have a finite $\e\to 0$ limit,
gives
\be 
\label{29}
Z[\gamma]= e^{iS_{1}[\gamma ]} \hs{1}
e^{-\frac{i}{8\pi G_{d+1}}\hs{1}\int_{\partial M} \sqrt{-\gamma}\hs{1}
(\frac{d-1}{l}
+ \frac{l}{2d-4} R + ... )} < e^{i \gamma_{\mu\nu}T^{\mu\nu}}>_{CFT}, 
\ee 
where $R$ is the Ricci scalar of $\gamma$ and we do not include
higher derivative terms. Redefining the metric on the boundary by
$g_{\mu\nu} = \frac{\e^{2}}{l^{2}} \gamma_{\mu\nu}$, (\ref{29})
becomes
\be \label{R}
Z=\int [dg] e^{-\frac{i}{16\pi
G_{d}}\hs{1} \int d^{d}x\hs{1} \sqrt{-g} \hs{1}\cal{R}+...}
\hs{1}<e^{ig_{\mu\nu}\tilde{T}^{\mu\nu}}>_{CFT}, 
\ee 
where $G_{d}$ is given by (\ref{eq2}) and the parameter $a$ is
again fixed by (\ref{a}). This shows that, the theory of gravity
with a negative cosmological constant defined by the partition
function (\ref{firstint}) in $d+1$ dimensions is, indeed,
equivalent to a $d$-dimensional gravity without a cosmological
constant coupled to a CFT on the boundary. Therefore, complementary
pictures arise since the theory defined by 
(\ref{firstint}) can be solved either by a semiclassical approximation 
giving a set of linearized modes or by first splitting the measure into 
bulk and boundary parts and then applying $AdS$/CFT correspondence.

\
\

It is straightforward to repeat the above considerations for a
minimally coupled massless scalar field on $AdS_{\e}$. Starting
from the action
\be
S_{s}=\int_{M}\hs{1}\sqrt{-G}\hs{1}\nabla_{A}\phi\hs{1}\nabla^{A}\phi ,
\ee
and allowing the scalar to vary on the boundary, one obtains 
\be
\Box \phi =0
\ee
\be
n^{A}\nabla_{A}\phi|_{\partial M}=0.
\ee
Therefore, in the compactification picture, perturbations around
the $\phi=0$ vacuum obey (\ref{diff})-(\ref{bound}), and this
gives rise to the same modes (\ref{soln}) which consists of a
single massless and a tower of continuum massive KK excitations.
The propagator obeys (\ref{green})-(\ref{greenb}) and thus, on
the boundary, can be written as (\ref{cgreen}).

\
\

In the $AdS$/CFT picture, the partition function 
\be
Z=\int [d\phi ]\hs{1}e^{iS_{s}},
\ee
where the sum is over all fields on $AdS_{\e}$, can be calculated as
\be
Z=\int [d\phi_{0}]\hs{1}[d\phi]_{\phi|_{\partial M}=\phi_{0}} \hs{1} e^{iS_{s}},
\ee
where the first sum is over all bulk fields which match a given
boundary field, and the second integral is over all boundary
fields. Using $AdS$/CFT this becomes
\be\label{scal}
Z=\int [d\phi_{0}]\hs{1} e^{i\int \hs{1}d^{d}x \hs{1}\frac{1}{2(d-2)}
(\phi_{0} \hs{1}\partial_{x}^{2}\hs{1}\phi_{0} + ....)}\hs{1} 
<e^{i\phi_{0}\hs{1}O}>_{CFT}, 
\ee
where $O$ is the dual CFT operator and the counterterms
have been calculated in \cite{skend}. The kinetic term in
(\ref{scal}) gives the standard $1/p^{2}$ propagator and the two
point function $<OO>_{CFT}$ gives rise to corrections.

\
\

It is easy to see that when $d=4$ the first order corrections to the
propagators calculated in two pictures agree with each other at large
distances. In the case of gravity this has been shown in \cite{gubser,
lin, duff}. In the compactification picture this correction is given by
$\Delta_{KK}$ in (\ref{KK}) for both gravity and scalar field. On the
other hand, in $AdS$/CFT picture, Ricci scalar in (\ref{R}) and the
kinetic term in (\ref{scal}) give the standard $d$-dimensional propagator
$1/p^{2}$. From (\ref{R}) and (\ref{scal}), the first order corrections to
this can be calculated to be $1/p^{2} <TT>_{CFT}1/p^{2}$ and
$1/p^{2}<OO>_{CFT}1/p^{2}$ for the metric and the scalar, respectively.
The $AdS$/CFT correspondence can independently be used to find
\be
<OO>_{CFT}\sim \frac{1}{x^{2d}}\sim <TT>_{CFT},
\ee
which shows that the first order corrections to the scalar and graviton
propagators are also equal to each other in the complementary $AdS$/CFT
picture. Therefore, the agreement found in the graviton propagator in two
pictures implies the same conclusion for the scalar field. 

\
\

In this letter we have showed that it is possible to reformulate
alternative to compactification scenario of \cite{sun} without introducing
singular brane sources. This has been achieved by replacing the brane
action in RS model with a term proportional to the area of the boundary of
the compactified AdS and letting the metric on the boundary vary. This
clarifies the relation between $AdS$/CFT duality and the RS scenario. The
approach presented in this paper also allows one to treat other fields
propagating in the bulk exactly like gravity. One important problem which
remains to be solved is the singularity of the modes on the horizon of
$AdS$. This problem also arises in the scenario of \cite{sun} and may
force one to introduce another boundary to cut off $z=\infty$ region.  The
consequences of such a modification remains to be explored.

\section*{Acknowledgments}
We are grateful to C.N. Pope for valuable discussions.

\end{document}